\documentclass[11pt]{article}

\usepackage[preprint]{acl}

\usepackage{times}
\usepackage{latexsym}

\usepackage[T1]{fontenc}

\usepackage[utf8]{inputenc}

\usepackage{microtype}

\usepackage{inconsolata}

\usepackage{graphicx}
\usepackage{xcolor}
\definecolor{lightorange}{HTML}{FFE6CC}  
\definecolor{lightblue}{HTML}{DAE8FC}    
\usepackage{algorithm}
\usepackage{algorithmic}
\usepackage{threeparttable}
\usepackage{changepage}
\usepackage{amsmath}
\usepackage{amssymb}

\usepackage{booktabs}
\usepackage{multirow}

\usepackage{url}

\usepackage{subcaption}
\usepackage{enumitem}
\usepackage{comment}
\graphicspath{{.}}

\setlist[itemize]{nosep,leftmargin=*}
\usepackage[font=small]{caption}
\title{From Regulatory Approvals to Patents: Cross-Domain Linking for Cardiovascular Device Traceability}

\author{\normalfont
  \begin{tabular}{p{0.21\textwidth} p{0.5\textwidth} p{0.21\textwidth}}
    \centering
    Qingqing Yang$^1$ \\
    $^1$HKUST (Guangzhou) \\
    Guangzhou, China
    &
    \centering
    Haijiang Liu$^{2,3}$\thanks{Co-corresponding authors: \texttt{bill1103478225@outlook.com};\texttt{moyanli@hkust-gz.edu.cn}.} \\
    $^2$Wuhan Univ. of Sci. \& Tech. \\
    $^3$Hubei Key Lab. of Intelligent 
    Info. Processing \& Real-time Industrial System \\
    Wuhan, China
    &
    \centering
    Moyan Li$^{1*}$ \\
    $^1$HKUST (Guangzhou) \\
    Guangzhou, China
    \\
  \end{tabular}
}

\begin{document}
\maketitle
\begin{abstract}\footnotetext{\url{https://github.com/myqqhub/Bridge-MedDevKG}.}
Linking FDA-approved medical devices to their underlying United States Patent and Trademark Office (USPTO) patents enables critical applications such as recall root-cause analysis, M\&A-driven IP discovery, and technology trajectory mapping. However, this cross-domain entity linking task remains unexplored due to severe \textit{semantic gaps}: FDA documents focus on clinical outcomes, while patents describe technical mechanisms, yielding minimal lexical overlap.
We formalize medical device-patent linking as a challenging cross-domain entity linking problem characterized by label scarcity and domain shifts. Using cardiovascular devices as a high-impact, representative domain featuring diverse technologies, high recall rates, and abundant disclosures, we construct a benchmark with 434 devices, 698K patents, and 585 high-fidelity expert-verified pairs.
To address these challenges, we propose Bridge-MedDevKG, a coarse-to-fine framework that integrates (1) \textbf{MedDevOnto}, a domain-specific ontology that anchors 
device concepts via three-tier UMLS normalization; (2) \textbf{Multi-signal candidate generation} fusing company affiliation, semantic similarity, and ontology-weighted entity overlap; and (3) \textbf{Heterogeneous reranking} with multi-signal scoring and XGBoost classification on hard negatives.
Our approach achieves a conservative lower-bound recall of 91.6\% on the gold standard with 50.9\% noise reduction, substantially outperforming LLM baselines under comparable evaluation. The resulting MedDevKG provides 6.8M high-confidence links, laying a scalable foundation for regulatory-IP integration across medical specialties.
\end{abstract}

\section{Introduction}
\begin{figure}[t] 
    \centering
    \includegraphics[width=\columnwidth]{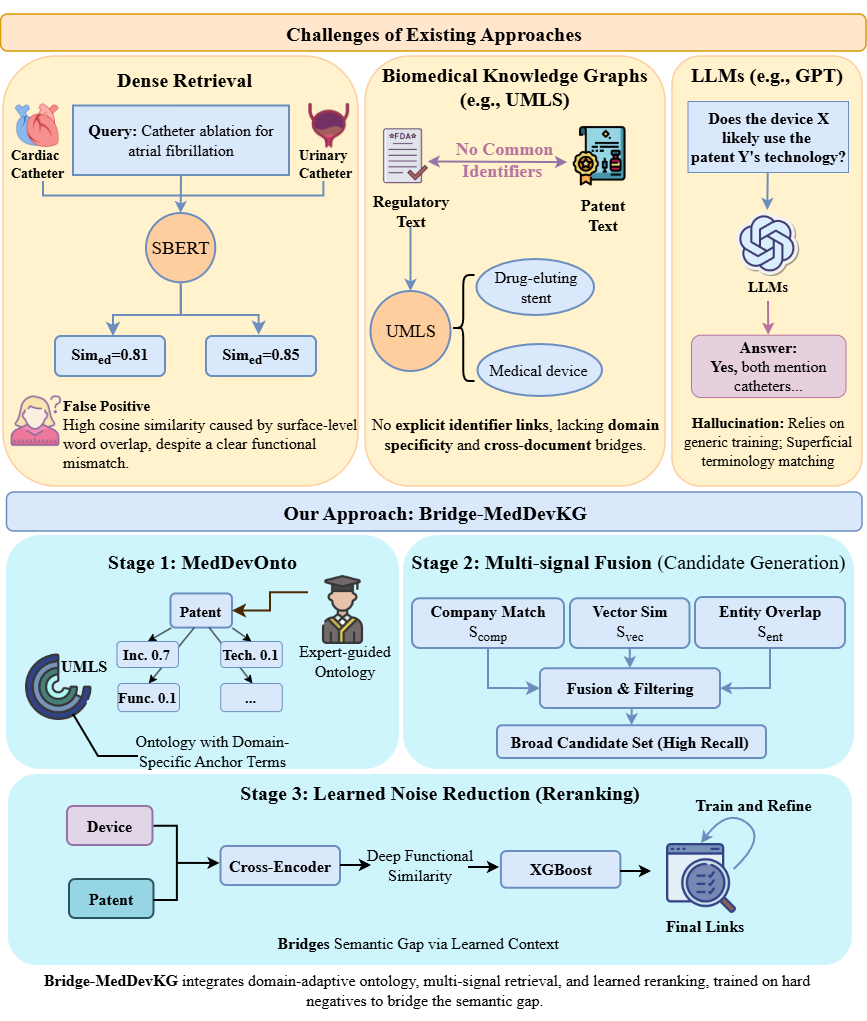}
    \caption{Comparison of approaches. (A) \colorbox{lightorange}{Existing methods} struggle with cross-domain ambiguity. (B) \colorbox{lightblue}{Our \textbf{Bridge-MedDevKG} } uses heterogeneous fusion to bridge the semantic gap where single-source methods fail.}
    \label{fig:comparison}
\end{figure}

Medical devices undergo a rigorous regulatory approval process before clinical deployment, while their underlying innovations are independently protected by their patents. These siloed regulatory and IP processes lack systematic cross-referencing. Automated linking would enable critical applications: \textit{recall root-cause analysis} by tracing recalled devices to implicated patent families, \textit{M\&A (mergers and acquisitions)-driven IP discovery} across corporate acquisitions, and \textit{technology trajectory mapping} for competitive intelligence.

This cross-domain entity linking task remains largely unexplored due to severe semantic gaps: the Food and Drug Administration (FDA) documents emphasize clinical outcomes (e.g., ``treats coronary artery disease''), while the United States Patent and Trademark Office (USPTO) patents describe technical mechanisms (e.g., ``helical coil with radiopaque markers''), yielding minimal lexical overlap (Jaccard similarity 0.039 on verified pairs).

In this work, we bridge the semantic gap via an \textbf{ontology with domain-specific anchor terms} and \textbf{multi-signal reranking}. Despite existing approaches providing valuable tools for cross-domain matching, such as dense retrieval methods \citep{karpukhin2020dense,reimers2019sentence}, biomedical knowledge graphs \citep{bodenreider2004unified,chandak2023building}, and Large Language Models (LLMs) matching and ranking \citep{sun2023chatgpt,qin2024large}, when applied to medical device-patent linking, they often show limitations like \textbf{conflating} semantically similar but functionally distinct entities (e.g., cardiac vs. urinary catheters), \textbf{struggling} to elevate domain-critical terms (e.g., ``stent'') over generic ones (e.g., ``device''), and \textbf{failing} to discriminate precise functional equivalence across regulatory and technical descriptions.

To address these challenges under realistic label scarcity, we propose \textsc{\textbf{Bridge-MedDevKG}}, a coarse-to-fine framework (Figure~\ref{fig:comparison}). It integrates \textsc{\textbf{MedDevOnto}}, an ontology with domain-specific anchor terms and three-tier UMLS concept normalization over the Unified Medical Language System (UMLS); \textbf{multi-signal candidate generation} fusing company affiliation, semantic similarity, and ontology-weighted entity overlap; and \textbf{heterogeneous reranking} with cross-encoder scoring and gradient-boosted classification on hard negatives.

Our contributions are:
\begin{enumerate}[leftmargin=*, nosep]
    \item The first formalization and benchmark for medical device-patent linking, with 585 high-fidelity expert-verified pairs under label scarcity.
    
    \item \textsc{MedDevOnto}, identifying device-critical anchor terms as discriminative matching pivots and enabling cross-document entity resolution through three-tier UMLS concept normalization.
    
    \item A label-efficient fusion pipeline achieving a conservative lower-bound recall of 91.6\% with 50.9\% noise reduction, outperforming LLM baselines under the reported evaluation settings.

    \item \textsc{MedDevKG}, linking 434 cardiovascular devices to 698K patents via 6.8M high-confidence relationships.
\end{enumerate}

\section{Related Work}
\paragraph{Cross-Domain Entity Linking.}
Entity linking (EL) connects mentions in text to canonical entries in a knowledge base. When source and target documents originate from different domains, the task becomes cross-domain entity linking, which presents unique challenges beyond standard EL \citep{shi2023knowledge} (formalized in \S\ref{sec:Task Challenges} as C1–C4). Recent advances include joint representation learning across domains \citep{soliman2022cross}, zero-shot transfer using BERT-based bi-encoders \citep{partalidou2022improving}, pattern exploitation for improved domain transfer in NER \citep{blair2022improving}, and pivot-based frameworks for cross-lingual settings \citep{rijhwani2019zero}. 

In our task, FDA approval documents describe clinical outcomes and safety profiles, whereas USPTO patents detail technical mechanisms and engineering specifications, resulting in fundamentally incompatible vocabularies that limit the ability of learned representations to bridge the gap.

\paragraph{Biomedical Knowledge Graphs.}
Structured knowledge resources have enabled significant advances in biomedical NLP. UMLS provides comprehensive cross-vocabulary normalization \citep{bodenreider2004unified}, widely adopted for named entity recognition and relation extraction \citep{yuan2021improving}. PrimeKG \citep{chandak2023building} integrates multiple resources for precision medicine applications, while PKG 2.0 \citep{xu2025pubmed} connects scientific papers, patents, and clinical trials through citation and grant linkages. Cross-terminology link mining \citep{patel2006mining} facilitates EHR integration via transitive relationship paths. In the patent domain, prior work extracts engineering knowledge graphs for design retrieval~\citep{zuo2022patent,siddharth2022engineering}, but focuses on \textit{within-patent} analysis rather than cross-domain regulatory-IP linking. 

Although these knowledge graphs capture structured knowledge of each field, they rely on \textit{explicit cross-references}: citations, shared author identifiers, or grant numbers, to establish connections between entities, whereas \textit{no standardized cross-reference exists between these two regulatory systems}. The relationship between a device and its underlying patents must therefore be inferred from indirect evidence such as company names, technical descriptions, and temporal patterns. Furthermore, standard entity matching typically assigns uniform importance to concepts within the same semantic type, failing to distinguish domain-critical terms (e.g., ``drug-eluting stent'') from generic ones (e.g., ``medical device'').\footnote{A concurrent study~\citep{cunningham2025linking} maps 
device \textit{classes} across CPC and CFR for ex-ante market 
exploration---a complementary but distinct task. We differ in 
granularity (instance-level device-to-patent linking for ex-post 
IP traceability), semantic gap (FDA clinical narratives vs.\ 
USPTO engineering claims; Jaccard\,=\,0.039), and output (a 
complete linkage for all FDA cardiovascular devices verified on 
a 585-pair expert benchmark).}

\paragraph{Retrieve-then-Rerank Architectures.}
Two-stage retrieve-then-rerank pipelines have become standard in information retrieval \citep{nogueira2019passage}. Dense retrieval with learned representations addresses vocabulary mismatch \citep{karpukhin2020dense}, while multi-aspect embeddings capture query-document relationships across multiple dimensions \citep{kong2022multi}. For domain adaptation, disentangled representations separate domain-invariant and domain-specific features \citep{zhan2022disentangled}, and regularized frameworks map heterogeneous domains onto shared latent spaces \citep{wang2009heterogeneous}. Structure-aware approaches integrate document relationships with semantic content \citep{raman2022structure}. 

These methods achieve strong results when queries and documents occupy related semantic spaces. But in our case, with severe label scarcity, where training examples are limited, and domains are fundamentally misaligned, challenges remain.

\paragraph{LLMs for Information Retrieval.}
Large language models have demonstrated impressive zero-shot capabilities on standard IR benchmarks. Listwise ranking with GPT-4 achieves competitive results on established test collections \citep{sun2023chatgpt}, progressive training strategies bridge language modeling objectives with ranking tasks \citep{zhang2023rankinggpt}, and uncertainty-aware approaches improve robustness \citep{zeng2024llm}. Pairwise ranking prompting has also shown strong zero-shot and few-shot performance on diverse retrieval tasks \citep{qin2024large}. 

However, when vocabulary disparity is severe, LLMs struggle to discriminate functional equivalence from surface similarity---a limitation particularly relevant for cross-domain matching where terminological conventions differ fundamentally.

Due to the nature of our task in terms of fundamentally different terminologies, indirect device-patent relationships inference, and distinguished terminology weighting for knowledge organization, we (1) build a domain-specific ontology with anchor terms and three-tier UMLS normalization that prioritizes clinically and technically critical concepts; (2) design a multi-signal fusion method combining structural, semantic, and knowledge-based evidence; and (3) construct a benchmark dataset of 585 expert-annotated device-patent pairs enabling systematic evaluation.

\section{Task Definition}
\label{sec:task}

We formalize \textit{medical device-patent linking} as a cross-domain entity linking task between the FDA-approved medical devices and the USPTO patents.

\subsection{Problem Formulation}

Let $\mathcal{D} = \{d_1, \dots, d_n\}$ be the set of the FDA Premarket Approval (PMA) summaries for cardiovascular devices and $\mathcal{P} = \{p_1, \dots, p_m\}$ the set of the USPTO patent abstracts in relevant medical classes. The goal is to predict a set of links $\mathcal{L} \subseteq \mathcal{D} \times \mathcal{P}$ such that $(d_i, p_j) \in \mathcal{L}$ if patent $p_j$ protects technology embodied in device $d_i$. This yields a highly asymmetric one-to-many mapping (median 5, maximum 83 patents per device).

\subsection{Data Construction and Benchmark}

We focus on cardiovascular devices, selected as a high-impact representative showcase of the broader task. This subdomain accounts for the majority of life-sustaining device approvals, exhibits the highest recall rates, spans diverse technologies (stents, catheters, valves, pacemakers, ablation systems, grafts), experiences frequent M\&A, and offers richer public patent disclosures than most other specialties, which enables reliable gold-standard construction while capturing the full spectrum of cross-domain challenges.

We construct the device corpus from 434 Class III cardiovascular PMA approvals (1976--2024), filtered by product codes and keywords, with manual exclusion of 29 non-cardiovascular entries (Appendix~\ref{app:data} for full criteria).

The patent corpus comprises 698,191 utility patents filtered by cardiovascular-relevant Cooperative Patent Classification (CPC) classifications (A61F2, A61M25, A61B5/6/8, etc.) with title keyword confirmation for manufacturing classes (Appendix~\ref{app:data}).

Company normalization uses a 29,758-entity dictionary covering subsidiaries and historical names to handle M\&A complexity (Appendix~\ref{app:company}).

The gold standard $\mathcal{G}$ consists of 585 expert-verified device-patent pairs sourced from public corporate disclosures (litigation, virtual patent marking, SEC filings), covering 88 devices (20.3\%). Table~\ref{tab:data_stats} summarizes key statistics.

\begin{table}[t]
\centering
\small
\caption{Dataset and benchmark statistics from data source (device, patent documents, and company entities) to gold standard patent-device relations.}
\label{tab:data_stats}
\begin{tabular}{lr}
\toprule
\textbf{Component}                  & \textbf{Count}          \\
\midrule
FDA Cardiovascular PMA Documents & 434         \\
USPTO Patents                     & 698,191     \\
Companies (normalized)        & 29,758 \\
Gold-Standard Verified Pairs      & 585         \\
Devices with Disclosures          & 88 (20.3\%) \\
Patents per Device (median/max)   & 5 / 83     
      \\
\bottomrule
\end{tabular}
\end{table}

\subsection{Task Challenges}
\label{sec:Task Challenges}
This task differs from standard entity linking in four aspects (Figure~\ref{fig:challenges}):

\begin{figure}[t]
    \centering
    \includegraphics[width=\columnwidth]{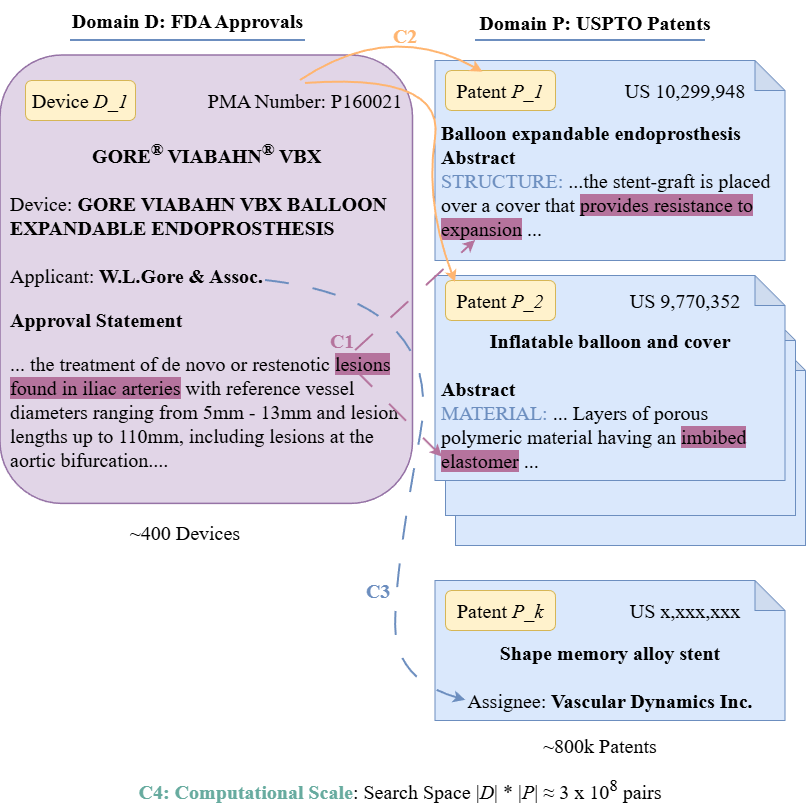}
    \setlength{\belowcaptionskip}{6pt}
    \caption{Overview of Cross-Domain Entity Linking challenges using GORE® VIABAHN® as a case study. Real Data Linkages: The device links to multiple patents (e.g., US 10,299,948 and US 9,770,352), demonstrating (C2) Granularity Asymmetry. The contrast between clinical indications ("lesions") and technical specifications ("resistance", "elastomer") highlights the (C1) Semantic Gap. Hypothetical Scenario: A fictional patent assignee is used to illustrate (C3) Corporate Structure challenges (M\&A resolution). (C4) Scale: Represents the massive pair-wise search space.}
    \vspace{-0.3cm}
    \label{fig:challenges}
\end{figure}

\textbf{C1: Semantic Gap.} FDA documents emphasize clinical outcomes (e.g., ``treats coronary artery disease'') while patents describe technical mechanisms (e.g., ``helical coil with radiopaque markers''). Lexical overlap is minimal: average Jaccard similarity between device and patent vocabulary is 0.039.\footnote{Computed over gold-standard pairs using unigram vocabularies after stopword removal.}

\textbf{C2: Granularity Asymmetry.} A single device integrates multiple patented technologies (median = 5, max = 83 patents per device), creating one-to-many alignment challenges.

\textbf{C3: Corporate Structure Complexity.} While only 10\% of verified links involve explicit assignee-manufacturer mismatches, frequent M\&A among major players requires cross-organization entity resolution.\footnote{E.g., Abbott acquired St.\ Jude Medical (2016) and CardioMEMS (2014).}

\textbf{C4: Scale and Label Scarcity.} With $|\mathcal{D}| \times |\mathcal{P}| \approx 3 \times 10^8$ candidate pairs, exhaustive evaluation is prohibitive, while expert-verified links cover only 88 devices (20.3\%).

\subsection{Evaluation Protocol}

Let \(\mathcal{D}\) and \(\mathcal{P}\) denote the sets of devices and patents, respectively. Let \(\mathcal{L}\) be the set of links returned by our pipeline. We use \(\mathcal{C}\) to denote the candidate pool size at each stage.

Given the partial nature of disclosures, we adopt conservative metrics:
(1) \textbf{Recall@Gold}: $|\mathcal{L} \cap \mathcal{G}| / |\mathcal{G}|$;
(2) \textbf{Noise Reduction}: $(|\mathcal{C}| - |\mathcal{L}|) / |\mathcal{C}|$;
(3) \textbf{FPR} (False Positive Rate): Reported specifically for LLM baseline comparisons.
\section{Methodology}
\label{sec:method}

\begin{figure*}[t]
    \centering
    \includegraphics[width=\linewidth]{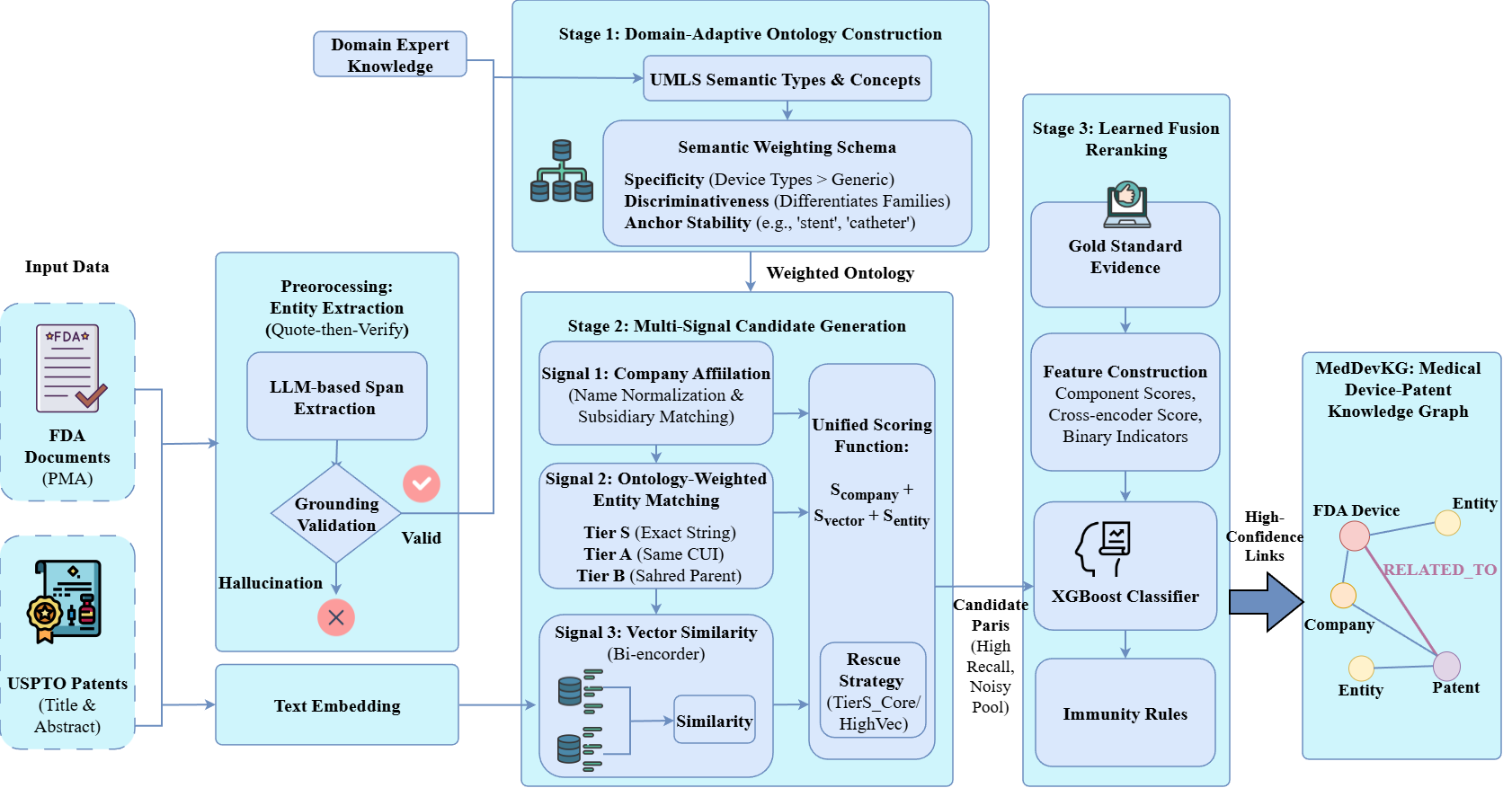}
    \caption{
    \textsc{Bridge-MedDevKG} pipeline: ontology with domain-specific anchor terms, multi-signal candidate generation, and learned reranking via cross-encoder + XGBoost.}
    \label{fig:pipeline}
\end{figure*}

We propose \textsc{Bridge-MedDevKG}, a coarse-to-fine framework that maximizes recall while progressively reducing noise (Figure~\ref{fig:pipeline}). Stages 1--2 construct a high-recall candidate pool using unsupervised and weakly supervised signals. Stage 3 applies learned reranking to bridge the semantic gap.

\subsection{Stage 1: MedDevOnto - Domain-Specific Ontology with Anchor Terms}
\label{sec:ontology}

UMLS \citep{bodenreider2004unified} provides comprehensive biomedical concept normalization but assigns uniform importance to all semantic types, treating generic terms (e.g., ``manufactured object'') equivalently to device-critical ones (e.g., ``stent''). Following middle-out ontology engineering \citep{uschold1996ontologies}, we construct \textsc{MedDevOnto} by differentially weighting UMLS semantic types according to three domain-specific  principles: 
\begin{enumerate}[noitemsep, topsep=0pt]
    \item \textbf{Specificity}: specific device types outweigh generic categories.
    \item \textbf{Discriminativeness}: terms distinguishing device families receive elevated weights.
    \item \textbf{Anchor Stability}: high-frequency terms in both corpora serve as matching anchors.
\end{enumerate}

Weights are assigned by UMLS semantic type and domain relevance:
\begin{itemize}[noitemsep, topsep=0pt]
    \item \textbf{High (1.0)}: Medical Device (T074), Therapeutic Procedure (T061), select diseases (T047).
    \item \textbf{Medium (0.5)}: Materials.
    \item \textbf{Low (0.1--0.3)}: Anatomy, generic manufactured objects.
    \item \textbf{Anchor terms} (stent, catheter, valve, etc.) receive weight 1.0 regardless of type.
\end{itemize}
 
We apply two-tier concept formalization: exact match after formalization (lowercasing, whitespace, and punctuation standardization), with fallback via syntactic head noun extraction, achieving 82.9\% entity-to-UMLS mapping coverage (Appendix~\ref{app:umls}).

Entity extraction and mapping use schema-constrained prompting with DeepSeek-V3, followed by Quote-then-Verify grounding (98.8\% fidelity; Appendix~\ref{app:extraction}). This yields three-tier matching: exact string, same UMLS Concept Unique Identifier (CUI), shared parent concept.

\subsection{Stage 2: Multi-Signal Candidate Generation}
\label{sec:candidate}

We fuse three complementary signals in a unified score $S(d,p)$:
\begin{equation}
    S(d, p) = S_{\text{company}} + S_{\text{vector}} + S_{\text{entity}}
\end{equation}
where $S_{\text{company}} \in \{0, 20\}$ is binary match using a 29,758-company dictionary; $S_{\text{vector}} \in [0, 65]$ is discretized Sentence-BERT cosine similarity; and $S_{\text{entity}}$ is ontology-weighted overlap across three tiers (Tier S: exact string, Tier A: same CUI, Tier B: shared parent).

Pairs enter the candidate pool if they achieve a composite score $S(d, p) \ge 70$, or satisfy rescue conditions\footnote{A rescue rule preserves below-threshold pairs when they contain strong single-signal evidence (Appendix \S\ref{sec:Rescue Strategy Contribution}).} for high-confidence single signals: Tier-S anchor match with core device terminology (weighted score $\ge 60$), or 
vector similarity $\ge 0.88$. Same-company pairs receive relaxed thresholds given the strong structural prior. All thresholds (company-match weight, composite score threshold, and rescue parameters) were selected on a held-out validation split comprising 20\% of the devices with gold-standard labels, and were \emph{not} tuned on the final evaluation set. This achieves 98.97\% gold recall at candidate generation (579/585 verified pairs) (Appendix~\ref{app:threshold}).

\subsection{Stage 3: Learned Noise Reduction}
\label{sec:rerank}
The candidate pool prioritizes recall but contains substantial noise. We apply heterogeneous feature fusion:

\textbf{Cross-Encoder Scoring}. We use BGE-M3 (BGE-reranker-v2-m3, 1024-token context) to compute deep contextual similarity:
\[
S_{\text{cross}}(d, p) = \text{CrossEncoder}([d; \texttt{SEP}; p])
\]

\textbf{XGBoost Classification}.
Each candidate is represented by a 9-dimensional feature vector including all Stage-2 scores, cross-encoder signal ($S_{\text{cross}}$),the raw SBERT cosine similarity ($\text{sim}_{\text{raw}}$), and binary indicators $\mathbf{b} \in \{0,1\}^3$. The binary vector $\mathbf{b} = [\mathbb{1}_{\text{core}}, \mathbb{1}_{\text{rescue}}, \mathbb{1}_{\text{org}}]$ captures structural and heuristic properties defined as:

$$ \mathbb{1}_{k} = \begin{cases} 1 & \text{if condition } k \text{ is satisfied} \\ 0 & \text{otherwise} \end{cases} $$
where $\mathbb{1}_{\text{core}}$ denotes an exact match on the core device term, $\mathbb{1}_{\text{rescue}}$ indicates the pair was admitted via the high-precision rescue rule, and $\mathbb{1}_{\text{org}}$ signifies that the device and patent share the same company assignment. 

The model is trained on gold positives and hard negatives (high-scoring non-gold pairs, $S \geq 70$, similarity $> 0.8$) with 5-fold cross-validation (F1 = 0.931).

\paragraph{Immunity Rules.}
High-confidence pairs (similarity $\geq 0.92$ with company match, or Tier-S anchor match) bypass classification to prevent false negatives (Appendix~\ref{app:immunity}).

\section{Experiments}
\label{sec:experiments}

\subsection{Experimental Setup}
We evaluate on the cardiovascular device-patent benchmark introduced in \S\ref{sec:task}: 434 FDA PMA documents, 698,191 USPTO patents, and a gold standard of 585 verified pairs across 88 devices (Table~\ref{tab:data_stats}).

\paragraph{Preprocessing.} Device and patent texts are used as-is (approval summaries and patent abstracts). Entity extraction for ontology matching employs schema-constrained prompting with DeepSeek-V3, achieving 98.8\% grounding fidelity via Quote-then-Verify (Appendix~\ref{app:extraction}). Company names are normalized using the 29,758-entity dictionary (Appendix~\ref{app:company}).

\paragraph{Implementation Details.}  
For sentence embeddings, we utilized \texttt{all-mpnet-base-v2}. The cross-encoder reranker employed is \texttt{BGE-M3}, chosen for its 1024-token context capability (refer to Appendix~\ref{app:reranker}).  XGBoost was implemented with 5-fold cross-validation, using gold positives alongside hard negatives; the probability threshold was optimized to achieve the maximum F1 score on validation, set at 0.22.  

All experiments were conducted on a single A800 GPU, with candidate generation being parallelized for enhanced efficiency.

\subsection{Baselines}
We compare against representative methods from each major paradigm:

\paragraph{Retrieval Baselines}  
We employ several retrieval baselines to evaluate the performance of the proposed methodologies. The lexical approaches utilized include Term Frequency-Inverse Document Frequency (\textbf{TF-IDF}) and \textbf{BM25} algorithms. For dense representations, we incorporate Sentence-BERT (\textbf{SBERT}) \citep{reimers2019sentence} with the \texttt{\textbf{all-mpnet-base-v2}} model, as well as \textbf{BioBERT} \citep{lee2020biobert} and \textbf{SapBERT} \citep{liu2021self}. Additionally, we implement a structural baseline that focuses on \textbf{company-only} matches following appropriate normalization procedures.

\paragraph{LLM Direct Classification}  
We prompt state-of-the-art models to classify device-patent relevance using extracted entities and functional descriptions (structured chain-of-thought prompting; full details in Appendix~\ref{app:llm}): GPT-4-turbo, GPT-4, Claude-3.5-Sonnet, Gemini-2.5-pro, DeepSeek-V3.

\paragraph{Cross-Encoder Baselines}  
We also include standalone rerankers that operate independently without the integration of multi-signal fusion or XGBoost methodologies. The models under consideration include \textbf{MiniLM-L12}, \textbf{TinyBERT-L6}, \textbf{BGE-M3}, as well as large language model-based rerankers such as \textbf{Qwen3-4B} and \textbf{ERank-4B}.

\section{Results}
\label{sec:results}

We investigate three evaluation questions:
\begin{itemize}[nosep, leftmargin=*]
    \item \textbf{EQ1}: How does our pipeline compare to retrieval and LLM baselines?
    \item \textbf{EQ2}: What drives the effectiveness of each component?
    \item \textbf{EQ3}: Does the framework generalize across device categories?
\end{itemize}

\subsection{EQ1: Main Performance Comparison}

Table~\ref{tab:main} presents results for retrieval baselines and LLM direct classification compared to our pipeline, respectively.

\begin{table*}[t]
\centering
\small
\caption{\textbf{Main Benchmarking Results.} 
Comprehensive comparison of retrieval baselines, zero-shot LLMs, and our Bridge-MedDevKG framework on 585 expert-verified device-patent pairs.
\textbf{R@Gold}: Recall on gold standard.
\textbf{FPR}: False positive rate on balanced evaluation set.
\textbf{Noise Red.}: Cumulative filtering ratio relative to full search space (300M pairs).
Best results in \textbf{bold}. Our framework substantially outperforms retrieval and LLM baselines in recall while enabling tractable full-corpus processing.}
\label{tab:main} 
\resizebox{\textwidth}{!}{%
\begin{tabular}{llcccccc}
\toprule
\textbf{Method Type} & \textbf{Model / Strategy} & \textbf{Links} & \textbf{R@10} & \textbf{R@100} & \textbf{R@Gold}$\uparrow$ & \textbf{FPR}$\downarrow$ & \textbf{Noise Red.}$\uparrow$ \\
\midrule
\multicolumn{8}{l}{\textit{\textbf{Group A: Retrieval Baselines} (50K patent subset)}} \\
Sparse & TF-IDF & --- & 5.6 & 16.5 & 35.5 & ---$^a$ & ---$^b$ \\
 & BM25 \citep{robertson2009probabilistic} & --- & 4.7 & 13.3 & 28.8 & ---$^a$ & ---$^b$ \\
Dense & BioBERT \citep{lee2020biobert} & --- & 3.9 & 18.3 & 35.6 & ---$^a$ & ---$^b$ \\
 & SapBERT \citep{liu2021self} & --- & 4.9 & 19.7 & 36.1 & ---$^a$ & ---$^b$ \\
 & SBERT \citep{reimers2019sentence} & --- & 8.2 & 24.0 & 39.3 & ---$^a$ & ---$^b$ \\
Structural & Company Name Matching & --- & 6.5 & 24.3 & 47.0 & ---$^a$ & ---$^b$ \\
\midrule
\multicolumn{8}{l}{\textit{\textbf{Group B: LLM Zero-shot Classifiers} (balanced evaluation set)}} \\
Commercial & Claude-3.5-Sonnet & --- & ---$^e$ & ---$^e$ & 29.6 & 11.0\% & ---$^c$ \\
 & Gemini-2.5-Pro & --- & ---$^e$ & ---$^e$ & 50.2 & 10.1\% & ---$^c$ \\
 & GPT-4 \citep{achiam2023gpt} & --- & ---$^e$ & ---$^e$ & 52.4 & 27.1\% & ---$^c$ \\
 & GPT-4-turbo \citep{achiam2023gpt} & --- & ---$^e$ & ---$^e$ & 60.1 & 28.6\% & ---$^c$ \\
Open Source & DeepSeek-V3  & --- & ---$^e$ & ---$^e$ & 23.5 & \textbf{9.5\%} & ---$^c$ \\
\midrule
\multicolumn{8}{l}{\textit{\textbf{Group C: Our Bridge-MedDevKG Framework} (full 698K corpus)}} \\
Baseline & All Candidates & $\sim$300M & ---$^f$ & ---$^f$ & 100.0 & --- & 0.0\% \\
Stage 2 & Candidate Generation & 13.9M & ---$^f$ & ---$^f$ & 98.97 & ---$^d$ & 95.4\% \\
Stage 3 & Full Pipeline$^\dagger$ & 6.8M & ---$^f$ & ---$^f$ & \textbf{91.61} & ---$^d$ & \textbf{97.7\%} \\
\bottomrule
\end{tabular}%
}

\vspace{1mm}
\begin{minipage}{\textwidth}
\footnotesize{
$^\dagger$Production configuration. Stage 3 achieves 50.9\% incremental noise reduction (13.9M $\to$ 6.8M).\\[1mm]
\textbf{Metric applicability:}
$^a$FPR requires explicit negative predictions; retrieval ranks without binary classification.
$^b$Noise reduction measures filtering efficiency; retrieval returns fixed top-K results.
$^c$LLMs evaluated on balanced sample, not full candidate space required for noise reduction.
$^d$Pipeline evaluated on full corpus; FPR requires same balanced set as LLM evaluation.
$^e$R@K metrics are for ranking tasks; LLMs perform binary classification without ranking.
$^f$R@K metrics are for ranking tasks; our pipeline outputs a filtered set, not a ranked list. \textbf{Design rationale:} 
Full-corpus evaluation for Group A requires 300M pairwise computations; Group B incurs prohibitive API costs ($>\$2,400$). Our coarse-to-fine architecture (Group C) is specifically designed to enable tractable full-corpus processing (Appendix~\ref{app:compute}). Groups B and C use the same 585 gold pairs as ground truth, but differ in evaluation protocol: LLMs are evaluated as binary classifiers on a balanced pair set, whereas our pipeline filters the full candidate space. For completeness, few-shot LLM results are reported in Appendix~\ref{app:llm}.
}
\end{minipage}
\end{table*}

\paragraph{Retrieval baselines struggle on cross-domain matching.}
No single retrieval method exceeds 50\% R@500 on the 50K patent subset (Table~\ref{tab:main}, Group A). General-domain SBERT (39.3\%) outperforms biomedical-specific models (BioBERT 35.6\%, SapBERT 36.1\%), suggesting that device documentation benefits from balanced engineering-clinical semantics rather than purely biomedical representations. Company matching achieves the highest single-signal performance (47.0\%), motivating its inclusion in multi-signal fusion. These results confirm a severe \textbf{vocabulary mismatch} between regulatory and technical documents.

\paragraph{LLMs exhibit strict recall-FPR tradeoff.}
Table~\ref{tab:main} (Group B) shows that no LLM\footnote{For completeness, few-shot LLM results are reported in Appendix~\ref{app:llm}} achieves both high recall and low FPR simultaneously. GPT-4-turbo reaches the highest recall (60.1\%) but with 28.6\% FPR---classifying nearly one-third of negative pairs as matches. DeepSeek-V3 minimizes FPR (9.5\%) but misses 76.5\% of valid links. This pattern reveals that LLMs match surface terminology (e.g., both documents mention ``catheter'') without discriminating functional specificity (cardiac ablation vs.\ urinary catheter).

\paragraph{Structured fusion substantially outperforms end-to-end approaches.}
Our full pipeline achieves a conservative lower-bound R@Gold of 91.6\% with 97.7\% cumulative noise reduction (Table~\ref{tab:main}, Group C). This represents a +31.5\% absolute improvement over the best LLM baseline (GPT-4-turbo). Table~\ref{tab:main} (Group C) details progressive filtering: Stages 1--2 reduce the search space by 95.4\% (from $\sim$300M to 13.9M candidates) while preserving 98.97\% gold recall; Stage 3 eliminates an additional 50.9\% of remaining noise, yielding 91.61\% final recall---a deliberate tradeoff favoring precision for downstream regulatory applications.

\subsection{EQ2: Component Analysis}
We analyze contributions at two levels: candidate construction (Stages 1--2) and learned reranking (Stage 3).

\paragraph{Candidate Construction Ablation.}
Table~\ref{tab:ablation_stage12} quantifies signal contributions. Vector similarity is most critical ($-$36.2\% when removed), providing the primary semantic bridge between regulatory and technical vocabularies. Entity matching contributes +11.3\%, supporting the value of device-critical anchor term identification and three-tier UMLS normalization over generic matching. Company matching adds +2.4\%, while the rescue strategy recovers +1.5\% edge cases with strong single-signal evidence.

\begin{table}[t]
\centering
\small
\caption{\textbf{Stage-2 core scoring ablation} 
Vector similarity dominates recall, while ontology-weighted entity overlap provides complementary gains; company and rescue mainly recover edge cases.}
\label{tab:ablation_stage12}
\begin{tabular}{p{0.5\linewidth}llc}
\toprule
\textbf{Configuration} & \textbf{R@Gold} & \textbf{$\Delta$} \\
\midrule
Core scoring & 70.77\% & --- \\
\quad w/o Company & 68.38\% & $-$2.39 \\
\quad w/o Vector & 34.53\% & $-$36.24 \\
\quad w/o Entity & 59.49\% & $-$11.28 \\
\quad w/o Rescue & 69.23\% & $-$1.54 \\
\bottomrule
\end{tabular}
\begin{tablenotes}
\footnotesize
\item[$\dagger$] $\theta$=70 + sim-rescue ($\mathrm{sim}\ge0.88$).
\end{tablenotes}
\end{table}

\paragraph{Learned Fusion vs.\ Rule-Based Thresholds.}
Table~\ref{tab:threshold} demonstrates why learned fusion outperforms fixed thresholds. Lowering the threshold to $\theta$=60 achieves near-perfect recall (99.8\%) but retains all noise. Raising to $\theta$=80 or $\theta$=90 reduces recall without proportional noise reduction (both plateau at 68.4\%). In contrast, Stage 3's learned approach achieves \textit{both} higher recall (+20.8\% over the Stage-2 core scoring configuration at $\theta$=70 with sim-rescue; Table~\ref{tab:ablation_stage12}) \textit{and} substantial noise reduction (50.9\%) by capturing non-linear signal interactions. For instance, a pair with moderate similarity (0.82) but strong company match and anchor term overlap may be valid, while one with high similarity (0.90) but no supporting evidence may be spurious---patterns that fixed thresholds cannot capture. Rule-based thresholds can also reduce noise, but only by sacrificing recall; Stage 3 improves this trade-off rather than replacing heuristic filtering altogether.

\begin{table}[t]
\centering
\small
\caption{Learned fusion vs.\ rule-based thresholds. Threshold baselines use score-only admission without rescue rules.\textit{Fixed thresholds fail to balance recall and noise, motivating learned fusion.}}
\label{tab:threshold}
\begin{tabular}{lcc}
\toprule
\textbf{Strategy} & \textbf{R@Gold} & \textbf{Noise Red.} \\
\midrule
Threshold $\theta$=60 & 99.83\% & 0\% \\
Threshold $\theta$=70 & 69.23\% & 0\%\\
Threshold $\theta$=80 & 68.38\% & 12.3\% \\
Threshold $\theta$=90 & 68.38\% & 28.7\% \\
\midrule
\textbf{Stage 3 (Learned)} & \textbf{91.6\%} & \textbf{50.9\%} \\
\bottomrule
\end{tabular}
\end{table}

\paragraph{Cross-Encoder Selection.}
Table~\ref{tab:reranker} compares cross-encoder architectures.\footnote{The net contribution of the cross-encoder to the full pipeline 
is a +1.03\% gain in gold recall (confirmed by ablating 
\texttt{ai\_score} under matched retraining) at an inference cost of 8.4\,ms per pair---a favourable cost--benefit trade-off given that cross-encoder scoring operates only on the reduced 13.9M candidate pool rather than the full 300M search space.} Here, \textit{ms} denotes average inference latency in milliseconds per pair. Traditional cross-encoders substantially outperform LLM-based rerankers (0.795 vs.\ 0.401 ROC-AUC). We deploy \textbf{BGE-M3}\footnote{Although TinyBERT-L6 achieves slightly higher PR-AUC, that comparison is based on shorter effective inputs and does not provide the same coverage for long, technically dense patents.} despite slightly lower ROC-AUC for two reasons: (1) higher PR-AUC (0.602 vs.\ 0.508), indicating superior positive-class performance in imbalanced settings; (2) 1024-token context window prevents truncation of long patent abstracts. All model differences are statistically significant ($p < 0.001$, bootstrap test).

\begin{table}[t]
\centering
\small
\begin{threeparttable}
\caption{Cross-encoder comparison.\textit{Traditional cross-encoders outperform LLM-based rerankers, with BGE-M3 offering the best trade-off between accuracy and context length.}}
\label{tab:reranker}
\begin{tabular}{p{0.35\linewidth}cccc}
\toprule
\textbf{Model} & \textbf{ROC} & \textbf{PR-AUC} & \textbf{F1} & \textbf{ms} \\
\midrule
MiniLM-L12 & \textbf{.795} & .508 & \textbf{.590} & 3.2 \\
TinyBERT-L6 & .772 & \textbf{.609} & .583 & 4.2 \\
BGE-M3$^\dagger$ & .760 & .602 & .574 & 8.4 \\
\midrule
Qwen3-4B & .401 & .203 & .402 & 41.1 \\
ERank-4B & .519 & .302 & .403 & 33.3 \\
\bottomrule
\end{tabular}
\begin{tablenotes}
\small
\item[$\dagger$] Deployed model (1024-token context).
\end{tablenotes}
\end{threeparttable}
\end{table}

\paragraph{XGBoost Feature Analysis.}
Five-fold cross-validation yields F1 = 0.931 $\pm$ 0.012 and ROC-AUC = 0.991 $\pm$ 0.004, confirming stable generalization. Feature importance analysis reveals \texttt{score\_total} (22.1\%), \texttt{score\_entity} (20.0\%), and \texttt{is\_same\_company} (18.2\%) as top contributors\footnote{Feature importance analysis reveals that semantic signals remain dominant: score\_total and score\_entity both exceed the company feature. This pattern suggests that the model is not driven primarily by company-structural signals from disclosed patents.}, while \texttt{ai\_score} contributes only 4.0\%. 
A direct feature ablation under matched retraining conditions confirms this: removing \texttt{ai\_score} reduces gold recall by 1.03 percentage points, indicating a refinement rather than primary role.

\paragraph{Entity Weighting Validation.}
Table~\ref{tab:weighting} validates our claim that \textit{which} entities to match matters more than \textit{how} to weight them. Once anchor terms are identified, specific numeric weights have a limited impact ($<$2\% variation). Gold pairs exhibit significantly higher weighted entity overlap than random pairs (mean 47.3 vs.\ 12.1, $p < 0.001$, Welch's $t$-test), confirming that entity overlap remains discriminative for filtering.

\begin{table}[t]
\centering
\small
\caption{Entity weighting sensitivity. \textit{We observe limited variation in recall across different entity weighting strategies when anchor entities are fixed, while their effects on filtering non-gold pairs are discussed in the text.}
}
\label{tab:weighting}
\begin{tabular}{lcc}
\toprule
\textbf{Strategy} & \textbf{R@Gold} & \textbf{$\Delta$} \\
\midrule
Expert-Guided (Core=60, Other=6) & 70.77\% & --- \\
Uniform High (All=60) & 70.94\% & +0.17 \\
Uniform Mid (All=30) & 70.94\% & +0.17 \\
Uniform Low (All=6) & 69.06\% & $-$1.71 \\
Binary (Existence Only) & 70.94\% & +0.17 \\
\bottomrule
\end{tabular}
\end{table}

\subsection{EQ3: Generalization Across Categories}

To verify robustness, we stratify the 88 devices with gold-standard links by primary clinical function. Table~\ref{tab:category} shows recall remains stable across functionally distinct categories (89.7--92.3\%), indicating that the anchor-term-based ontology layer and multi-signal fusion generalize without overfitting to specific device types.

\begin{table}[t]
\centering
\small
\caption{Performance by device category. \textit{Recall remains stable across major cardiovascular device types.}}
\label{tab:category}
\begin{tabular}{lcc}
\toprule
\textbf{Category} & \textbf{Gold Pairs} & \textbf{R@Gold} \\
\midrule
Stents & 142 & 92.3\% \\
Catheters & 98 & 90.8\% \\
Valves & 65 & 91.5\% \\
Pacemakers/ICDs & 54 & 89.7\% \\
\bottomrule
\end{tabular}
\end{table}

\section{Conclusion}
\label{sec:conclusion}

We presented \textbf{Bridge-MedDevKG}, a framework for cross-domain entity linking between FDA-approved medical devices and USPTO patents. Our contributions include: (1) the first formalization and benchmark for device-patent linking with expert-verified evaluation pairs; (2) \textbf{MedDevOnto}, which identifies device-critical anchor terms and enables three-tier UMLS-based cross-document entity resolution; and (3) a heterogeneous signal fusion pipeline that significantly outperforms LLM baselines in both recall and noise reduction. Our experiments demonstrate that LLM direct classification struggles in zero-shot and few-shot settings on this task, validating the need for structured multi-signal approaches.

The resulting \textbf{MedDevKG} enables practical applications in recall surveillance, M\&A IP discovery, and technology trajectory analysis (see Appendix~\ref{app:cases} for case studies).
\section*{Limitations}

\paragraph{Evaluation Protocol Constraints.}
Our evaluation relies on corporate disclosures (litigation records, SEC filings) as ground truth. Despite exhaustive manual search across all FDA PMA-approved cardiovascular devices, we could only identify 585 verifiable device-patent pairs for 88 devices. This scarcity reflects the inherent opacity of medical device IP landscapes: companies strategically disclose only a fraction of relevant patents. Consequently, pairs absent from the gold standard may still be valid associations, making our recall estimates conservative lower bounds. This also renders standard precision metrics less interpretable—predicted links not in the gold standard are not necessarily false positives. 

\paragraph{Benchmark Scale and Scope.}
Our gold standard of 88 devices with 585 pairs, though it represents a reasonably comprehensive publicly verifiable collection for this task, remains limited for fine-grained statistical analysis. The 18 anchor terms and UMLS semantic type weights are calibrated specifically for cardiovascular devices; other therapeutic areas (e.g., orthopedics, neurology) exhibit distinct lexical patterns and may require domain-specific recalibration of both ontology weights and signal fusion parameters.\footnote{The core pipeline remains largely device-agnostic, including multi-stage candidate generation, multi-signal fusion, XGBoost reranking, and immunity-style filtering. Domain transfer mainly requires recalibration of a compact peripheral layer, including anchor terms and score thresholds, for specialties such as orthopedics or neurology.} 

\paragraph{Methodological Design Choices.}
Our framework incorporates heuristic components (similarity thresholds, immunity rules, rescue strategies) alongside learned models. While these pragmatic choices optimize real-world performance, they introduce hyperparameters requiring tuning for new domains. Our Stage 3 reranker is designed primarily for noise reduction rather than recall improvement; the modest contribution of cross-encoder scores (4.0\% feature importance) relative to structural signals reflects this design intent—semantic similarity serves as a refinement signal atop robust entity-matching features. Additionally, our ablation on ontology weighting (Table 6) suggests that anchor term identification may matter more than specific weight assignments; minimal-supervision alternatives could improve transferability. Our sensitivity analysis and per-category results (Table~\ref{tab:category}; 
recall 89.7--92.3\% across stents, catheters, valves, and pacemakers/ICDs) suggest that this heuristic layer is compact rather than brittle within the cardiovascular domain; nevertheless, cross-specialty validation remains future 
work.

\paragraph{Recall-Noise Trade-off.}
Our Stage 3 reranking trades 7.4\% recall for 50.9\% noise reduction, reflecting an explicit design choice. Applications requiring exhaustive coverage should use Stage 2 outputs directly (98.97\% recall), accepting higher noise in exchange for near-complete retrieval.

\paragraph{Future Directions.}
Key extensions include: (1) refined evaluation metrics—among the 49 unrecovered gold pairs, expert audit reveals that 60.9\% are weak associations (tangentially related patents disclosed for legal completeness), 30.4\% are genuine semantic failures, and 8.7\% reflect questionable gold labels; future work should develop stratified evaluation protocols that distinguish association strength levels and incorporate expert-annotated link quality scores, enabling more accurate assessment of true system recall; (2) multilingual and international expansion—our current focus on English USPTO-FDA documents leaves international regulatory-IP linking unexplored; extending to multilingual patent corpora (EPO, CNIPA, JPO) and harmonizing cross-jurisdictional device classifications would enable global IP landscape analysis; (3) cross-domain validation on orthopedics and neurology devices; (4) precision estimation via expert annotation of sampled predictions; (5) temporal M\&A modeling for dynamic corporate genealogies; and (6) part-whole compositional reasoning for multi-component devices.

\section*{Ethics Statement}

This research adheres to ethical standards in biomedical informatics and knowledge graph construction. All datasets are derived from publicly available, appropriately licensed sources: FDA Premarket Approval (PMA) approval statements and USPTO patent records, both accessible through official government databases. No personally identifiable information is involved, as our analysis focuses exclusively on institutional entities (companies, regulatory bodies) and technical documentation.

We develop scalable tools for cross-domain entity linking to promote transparency in the medical device ecosystem. Through systematic device-patent mapping, we aim to facilitate recall root-cause analysis, technology trajectory tracking, and intellectual property landscape understanding---applications that benefit patients, researchers, and regulatory bodies alike.

We recognize several limitations and potential risks. First, our gold standard is constructed from corporate disclosures, which may reflect strategic rather than comprehensive patent associations; thus, pairs absent from the gold standard should not be assumed incorrect. Second, the knowledge graph captures correlational relationships between devices and patents, not causal or legal dependencies---it should complement, not replace, expert judgment in regulatory or litigation contexts. Third, automated entity extraction and linking may propagate errors from source documents or introduce biases from the underlying language models.

The resulting \textsc{MedDevKG} is intended for research and analytical purposes. We caution against using it as the sole basis for legal, regulatory, or investment decisions without independent verification. Future work should incorporate additional validation mechanisms and broader stakeholder input to enhance reliability and fairness.


\bibliography{custom}

\appendix

\section{Data Collection and Filtering}
\label{app:data}

\subsection{Patent Corpus Construction}

We construct the patent corpus from USPTO PatentsView (1976--October 2024) through multi-stage filtering:

\paragraph{Step 1: Patent Type.} Retain utility and reissue patents; exclude design patents lacking technical claims.

\paragraph{Step 2: Assignee Type.} Retain corporate / institutional assignees (USPTO codes 2, 3, 6--15); exclude individual inventors to focus on commercially relevant IP.

\paragraph{Step 3: CPC Classification.} We filter patents using a domain-specific schema of CPC prefixes defined in Table~\ref{tab:cpc_full}. To ensure precision, candidates falling under broad Manufacturing categories (e.g., B23P, C25D) are retained only if their titles contain cardiovascular-specific keywords (e.g., ``stent,'' ``catheter,'' or ``valve'') matching our regex constraints.

\begin{table}[h]
\centering
\small
\caption{Complete CPC classification schema for cardiovascular patents.}
\label{tab:cpc_full}
\resizebox{\linewidth}{!}{
\begin{threeparttable}
\begin{tabular}{lp{5.5cm}}
\toprule
\textbf{Category} & \textbf{CPC Codes} \\
\midrule
Core Prostheses & A61F2 (e.g., A61F2/82 stents, A61F2/24 valves) \\
Catheters & A61M25 (e.g., A61M25/10 balloons) \\
Diagnostics & A61B5, A61B6, A61B8, A61B17, A61B34 \\ 
Materials & A61L31, A61L27 \\
Manufacturing$^{\ast}$ & B23P15, B21D53, C25D5, C23C14, C22C38\\
\bottomrule
\end{tabular}
\begin{tablenotes}
        \footnotesize
        \item[*] Requires keyword matching (regex) to exclude general industrial patents.
    \end{tablenotes}
\end{threeparttable}
}
\end{table}

\subsection{Device Corpus Construction}

\paragraph{Cardiovascular Filtering.} We apply dual-layer filtering to FDA PMA approvals:
\begin{enumerate}[nosep]
    \item \textbf{Keyword matching}: Device name or trade name contains cardiovascular terms: \textit{cardio, vascular, coronary, atrial, heart, stent, valve, artery, aortic, mitral, pacemaker, defibrillator, ablation, angioplasty, graft, catheter, atherectomy, embolectomy, oximeter, electrode, annuloplasty, cannula, occluder, arrhythmia}
    \item \textbf{Product code matching}: FDA product codes in cardiovascular categories (DXY, LWS, NKE, PAQ, NPT, NIQ, MIH, LJP, MIP, MAJ, etc.)
\end{enumerate}

\paragraph{Manual Exclusion.} We curate an exclusion list of 29 non-cardiovascular devices that passed initial filters (Table~\ref{tab:exclusion}).

\begin{table}[h]
\centering
\small
\caption{Non-cardiovascular device exclusions by category.}
\label{tab:exclusion}
\resizebox{\columnwidth}{!}{%
    \begin{tabular}{lrl}
    \toprule
    \textbf{Category} & \textbf{Count} & \textbf{Example PMA} \\
    \midrule
    Orthopedic/Spine & 13 & P000028 (Cervical Cage) \\
    Neuro/Spinal & 3 & P810033 (Epidural Electrode) \\
    Gynecology & 4 & P010013 (Endometrial Ablation) \\
    Dental & 4 & P040013 (Dental Bone Graft) \\
    Ophthalmology & 4 & P080030 (Glaucoma Implant) \\
    Other & 1 & P010020 (Fecal Incontinence) \\
    \bottomrule
    \end{tabular}%
}
\end{table}

\subsection{Final Statistics}

After filtering: 698,191 patents from 11.2M USPTO records (6.2\% retention); 434 FDA PMA cardiovascular devices after excluding 29 non-cardiovascular entries.

\section{Company Normalization}
\label{app:company}

Medical device industry M\&A creates complex corporate genealogies that complicate assignee-manufacturer matching. We construct a normalization dictionary through three sources:

\paragraph{SEC Filings.} Extract subsidiary relationships from 10-K annual reports of major device manufacturers (Medtronic, Abbott, Boston Scientific, Johnson \& Johnson, Edwards Lifesciences, etc.).

\paragraph{Historical Records.} Track corporate name changes and acquisitions. Key examples:
\begin{itemize}[nosep, leftmargin=*]
    \item Abbott $\leftarrow$ St.\ Jude Medical (2016, \$25B)
    \item Abbott $\leftarrow$ CardioMEMS (2014)
    \item Medtronic $\leftarrow$ Covidien (2015, \$50B)
    \item Boston Scientific $\leftarrow$ Guidant (2006, \$27B)
    \item Endologix $\leftarrow$ TriVascular (2016, \$211M)
\end{itemize}

\paragraph{Abbreviation Expansion.} Map common abbreviations and trading names:
\begin{itemize}[nosep, leftmargin=*]
    \item J\&J $\to$ Johnson \& Johnson
    \item BSC $\to$ Boston Scientific Corporation
    \item MDT $\to$ Medtronic plc
\end{itemize}

\paragraph{Statistics.} The resulting dictionary covers 29,758 companies. Among these, 84 are \textit{bridging companies}---organizations/companies with both USPTO patents and FDA device approvals---enabling direct manufacturer-assignee linkage.

\paragraph{Commercial Reference Data.} We also use Compustat-based standardization records to reconcile historical naming variants across public-company filings. Because Compustat is a proprietary commercial database, the full normalization dictionary cannot be released publicly.

\section{Entity Extraction Details}
\label{app:extraction}

\subsection{Extraction Schema}

We define a rigorous schema to bridge the linguistic gap between regulatory and technical documents. Unlike standard named entity recognition (NER), our schema explicitly incorporates software and digital functions, reflecting the complexity of modern cardiovascular devices:

\begin{itemize}[nosep, leftmargin=*]
    \item \textbf{\textsc{Component} (Hardware \& Software)}:
    \begin{itemize}[nosep]
        \item \textit{Physical}: Stents, catheters, balloons, sensors, valves.
        \item \textit{Digital}: AI algorithms, mapping software, control units, user interfaces.
    \end{itemize}
    \item \textbf{\textsc{Mechanism} (Clinical \& Digital)}:
    \begin{itemize}[nosep]
        \item \textit{Clinical}: Ablation, angioplasty, pacing, hemostasis.
        \item \textit{Functional}: Signal analysis, image reconstruction, remote monitoring.
    \end{itemize}
\end{itemize}

\subsection{Prompting Strategy}

We employ \textbf{Schema-Guided Instruction Prompting} with DeepSeek-V3. Rather than relying solely on few-shot examples, we construct a structured system prompt containing three critical constraint layers to ensure standardization at the source:

\paragraph{1. In-Context Standardization.}
The model is instructed to perform implicit normalization during extraction. For instance, engineering descriptions like ``expandable prosthesis'' are mapped to ``Stent,'' and ``RF thermal heating'' is mapped to ``Cardiac Ablation'' (as seen in the \textit{Patent Translation Rules} section of our prompt).

\paragraph{2. The Anti-Super-Node Rule.}
To prevent the knowledge graph from degenerating into generic nodes, we enforce a specificity constraint: generic terms (e.g., ``System,'' ``Device,'' ``Method'') are forbidden unless modified (e.g., ``Stent Delivery System''). This directly addresses the granularity challenge (\S\ref{sec:task}, C2).

\paragraph{3. Domain-Specific Role Play.}
We use distinct system personas:
\begin{itemize}[nosep, leftmargin=*]
    \item \textbf{FDA Extractor}: Acts as a ``Precision Extractor'' focusing on identifying clinical indications and standardized product codes.
    \item \textbf{Patent Decoder}: Acts as a ``Translator'' to convert obfuscated engineering embodiments into standardized clinical functional terms.
\end{itemize}

\subsection{Quote-then-Verify Validation}

To ensure extraction fidelity, each extracted span must be grounded in the source document. We apply multi-tier matching:

\begin{enumerate}[nosep]
    \item \textbf{Exact match}: Span appears verbatim in source.
    \item \textbf{Case-insensitive}: Span matches after lowercasing.
    \item \textbf{Fuzzy match}: Levenshtein similarity $\geq$ 0.85.
\end{enumerate}

Non-grounded spans are rejected. Table~\ref{tab:validation} summarizes validation results.

\begin{table}[h]
\centering
\small
\caption{Quote-then-Verify validation statistics.}
\label{tab:validation}
\begin{tabular}{lrr}
\toprule
\textbf{Match Type} & \textbf{Count} & \textbf{Percentage} \\
\midrule
Exact match & 700,054 & 97.9\% \\
Case-insensitive & 1,686 & 0.2\% \\
Fuzzy match & 4,785 & 0.7\% \\
Rejected (hallucination) & 8,383 & 1.2\% \\
\midrule
\textbf{Total validated} & \textbf{706,525} & \textbf{98.8\%} \\
\bottomrule
\end{tabular}
\end{table}

\subsection{Final Entity Statistics}

The extraction process yields:
\begin{itemize}[nosep, leftmargin=*]
    \item 21,002 unique \textsc{Component} entities
    \item 12,057 unique \textsc{Mechanism} entities
    \item 714,907 total entity mentions across all documents
    \item 384-dimensional Sentence-BERT embeddings for all documents
\end{itemize}

\section{UMLS Mapping Details}
\label{app:umls}

\subsection{Mapping Procedure}

We map extracted entities to UMLS concepts using a two-tier strategy:

\paragraph{Tier 1: Exact Match.} After surface normalization (lowercasing, whitespace standardization, punctuation removal), we match against 2.35M UMLS terms.

\paragraph{Tier 2: Head Noun Extraction.} For multi-word terms failing exact match, we extract the syntactic head noun using spaCy and attempt matching. Example: ``bi-directional steerable catheter'' $\to$ ``catheter'' $\to$ C0085590.

\subsection{Coverage Statistics}

\begin{table}[h]
\centering
\small
\caption{UMLS mapping statistics.}
\label{tab:umls_stats}
\begin{tabular}{lrr}
\toprule
\textbf{Match Tier} & \textbf{Count} & \textbf{Percentage} \\
\midrule
Tier 1 (Exact) & 2,999 & 9.2\% \\
Tier 2 (Head Noun) & 24,096 & 73.7\% \\
No Match & 5,580 & 17.1\% \\
\midrule
\textbf{Total Coverage} & \textbf{27,095} & \textbf{82.9\%} \\
\bottomrule
\end{tabular}
\end{table}

\subsection{Semantic Type Distribution}

Table~\ref{tab:tui_dist} shows the distribution of mapped concepts across UMLS semantic types (TUIs).

\begin{table}[h]
\centering
\small
\caption{Top 10 UMLS semantic types in mapped entities.}
\label{tab:tui_dist}
\begin{tabular}{llr}
\toprule
\textbf{TUI} & \textbf{Semantic Type} & \textbf{Count} \\
\midrule
T073 & Manufactured Object & 5,913 \\
T074 & Medical Device & 5,899 \\
T169 & Functional Concept & 3,355 \\
T061 & Therapeutic/Preventive Procedure & 2,170 \\
T170 & Intellectual Product & 1,690 \\
T058 & Health Care Activity & 1,410 \\
T080 & Qualitative Concept & 894 \\
T059 & Laboratory Procedure & 835 \\
T081 & Quantitative Concept & 758 \\
T060 & Diagnostic Procedure & 663 \\
\bottomrule
\end{tabular}
\end{table}

\section{LLM Classification Study}
\label{app:llm}

\subsection{Experimental Setup}

\paragraph{Sampling.} From 85 devices with gold-standard links, we construct an evaluation set stratified by:\footnote{Three of the 88 disclosed devices lacked complete document representations required for structured prompt construction.}

\begin{itemize}[nosep, leftmargin=*]
    \item \textbf{Gold positives}: Verified device-patent pairs from corporate disclosures
    \item \textbf{Hard negatives}: High-similarity pairs ($\text{sim} > 0.8$) not in gold standard
    \item \textbf{Random negatives}: Randomly sampled non-gold pairs
\end{itemize}

Hard negatives are sampled from high-scoring non-gold pairs ($S \geq 70$, $\text{sim} > 0.8$) to challenge the classifier with difficult cases.

Entities failing Quote-then-Verify validation (1.2\%) are discarded before downstream matching.

\paragraph{Prompt Engineering.} We iteratively refine prompts through three strategies:
\begin{enumerate}[nosep]
    \item Direct yes/no classification
    \item Chain-of-thought reasoning
    \item Structured criteria for functional matching
\end{enumerate}
All models use the final structured prompt requiring explicit reasoning about component overlap, functional alignment, and temporal plausibility.

\paragraph{Models Evaluated.} GPT-4, GPT-4-turbo, Claude-3.5-Sonnet, Gemini-2.5-pro, Gemini-2.5-flash, Qwen2.5-72B, Qwen3-32B, GLM-4-32B, Kimi-K2, DeepSeek-V3.

\subsection{Full Results}

We first evaluate few-shot prompting effectiveness on a held-out balanced test set (30 gold positive pairs and 30 hard/random negative pairs, n=60, Table~\ref{tab:fewshot_llm}). 

\begin{table}[h]
\centering
\caption{Few-shot classification performance (Recall / FPR) on the held-out balanced sample.}
\label{tab:fewshot_llm}
\resizebox{\linewidth}{!}{
\begin{tabular}{lccc}
\toprule
\textbf{Model}          & \textbf{0-shot}   & \textbf{1-shot}    & \textbf{3-shot}    \\
\midrule
GPT-4-turbo             & 56.7 / 6.7        & 66.7 / 10.0        & 66.7 / 13.3        \\
DeepSeek-V3             & 40.0 / 0.0        & 53.3 / 6.7         & 66.7 / 6.7         \\
\bottomrule
\end{tabular}
}
\end{table}

Our pipeline achieves \textbf{91.6\% recall} on the full gold-standard set. Even the best few-shot LLM result (66.7\%) remains 24.9 percentage points below our pipeline. Note that 0-shot figures differ slightly from those in Table~\ref{tab:main} due to the smaller sample size (n=60), but the performance gap remains consistent. This confirms that the gap primarily stems from severe cross-domain vocabulary mismatch (Jaccard similarity = 0.039) rather than suboptimal prompting.

Table~\ref{tab:llm_full} presents complete results (0-shot) across all evaluated models.

\begin{table}[h]
\centering
\small
\caption{LLM direct classification results (0-shot, 85 devices).}
\label{tab:llm_full}
\resizebox{\linewidth}{!}{
\begin{tabular}{lcccc}
\toprule
\textbf{Model} & \textbf{Recall (\%)} & \textbf{Error (\%)} & \textbf{FPR (\%)} \\
\midrule
GPT-4-turbo & 60.1 & 39.9  & 28.6 \\
GPT-4 & 52.4 & 47.6  & 27.1 \\
Kimi-K2 & 54.7 & 45.3  & 20.5 \\
Gemini-2.5-pro & 50.2 & 49.8  & 10.1 \\
Gemini-2.5-flash & 45.8 & 54.2  & 12.3 \\
Qwen2.5-72B & 42.1 & 57.9 & 15.7 \\
Qwen3-32B & 38.9 & 61.1  & 14.2 \\
GLM-4-32B & 35.6 & 64.4  & 13.8 \\
Claude-3.5-Sonnet & 29.6 & 58.5 & 11.0 \\
DeepSeek-V3 & 23.5 & 76.5 & 9.5 \\
\bottomrule
\end{tabular}
}
\end{table}

\subsection{Error Analysis}

Models exhibit a clear recall-FPR tradeoff: aggressive models (GPT-4-turbo) achieve higher recall but suffer from excessive false positives; conservative models (DeepSeek-V3) minimize FPR but miss the majority of valid links.

Qualitative analysis reveals that LLMs rely heavily on surface terminology matching---e.g., identifying ``catheter'' in both documents---without discriminating functional specificity. A ``cardiac ablation catheter'' and ``urinary catheter'' may both be classified as related to a catheter patent, despite entirely different clinical applications.

\section{Immunity Rules Details}
\label{app:immunity}

Immunity rules prevent false negatives on high-confidence matches that the XGBoost classifier might incorrectly reject due to threshold conservatism:

\begin{itemize}[nosep]
    \item \textbf{High-similarity with company match} (sim $\geq 0.92$, same company): Strong semantic alignment confirmed by organizational evidence makes false positives unlikely.
    \item \textbf{Tier-S anchor match}: Exact string match on domain-critical anchor terms indicates precise functional correspondence.
\end{itemize}

Ablation (Table~\ref{tab:threshold}) shows removing immunity rules drops recall from 91.6\% to 85.27\% ($-$6.3\%) while only improving noise reduction by 4.4\%.
\section{Threshold Selection}
\label{app:threshold}
\subsection{Gold Pair Similarity Distribution}

We analyze the similarity distribution of 585 expert-verified 
device-patent pairs (Table~\ref{tab:sim_dist}). The distribution 
is concentrated in the high-similarity region, with all verified 
pairs exceeding 0.83 cosine similarity.

\begin{table}[h]
\centering
\small
\caption{Gold pair similarity distribution.}
\label{tab:sim_dist}
\begin{tabular}{lc}
\toprule
\textbf{Statistic} & \textbf{Value} \\
\midrule
Mean & 0.885 \\
Std & 0.021 \\
Min / Max & 0.832 / 0.940 \\
\midrule
50th percentile & 0.885 \\
75th percentile & 0.898 \\
95th percentile & 0.924 \\
\bottomrule
\end{tabular}
\end{table}

\subsection{Sensitivity Analysis of Stage-2 Heuristics}
\label{sec:stage2_sensitivity}

We examine the sensitivity of Stage-2 candidate generation to the composite
score threshold $\theta$ without rescue rules.
Table~\ref{tab:score_sens} shows that lower thresholds ($\theta \le 65$)
retain near-complete gold recall but admit substantially more candidates,
whereas stricter thresholds sharply reduce recall.
We therefore select $\theta=70$ as a practical operating point for Stage 2,
to be complemented by the rescue strategy described below.

\begin{table}[h]
\centering
\small
\caption{Score threshold sensitivity (no rescue).}
\label{tab:score_sens}
\begin{tabular}{ccc}
\toprule
\textbf{$\theta$} & \textbf{Hits} & \textbf{Recall} \\
\midrule
60 & 584 & 99.83\% \\
65 & 584 & 99.83\% \\
\textbf{70} & \textbf{405} & \textbf{69.23\%} \\
75 & 383 & 65.47\% \\
80 & 383 & 65.47\% \\
90 & 237 & 40.51\% \\
\bottomrule
\end{tabular}
\end{table}

\subsection{Rescue Strategy Contribution}
\label{sec:Rescue Strategy Contribution}
To recover valid pairs with weak composite scores but strong individual signals, we apply layered rescue rules (Table~\ref{tab:rescue}). This design improves recall without relaxing the global admission threshold for all pairs. Using sim $\ge 0.83$ directly would substantially expand the candidate set and introduce excessive noise, making downstream reranking intractable.

\begin{table}[h]
\centering
\small
\caption{Rescue rule contribution ($\theta=70$).}
\resizebox{\linewidth}{!}{
\label{tab:rescue}
\begin{tabular}{lcc}
\toprule
\textbf{Configuration} & \textbf{Recall} & \textbf{$\Delta$} \\
\midrule
$\theta \ge 70$ only & 69.23\% & --- \\
+ Entity $\ge 60$ (Core) & 69.23\% & +0.00\% \\
+ sim $\ge 0.88$ & 70.77\% & +1.54\% \\
+ sim $\ge 0.83$ (oracle lower bound) & 100.00\% & +29.23\% \\
\bottomrule
\end{tabular}
}
\end{table}

The empirical lower bound of 0.83 is justified by the gold-pair distribution: 100\% of verified pairs have similarity above this value (min=0.832). In practice, we apply a strict rescue threshold of 0.88, while same-company pairs receive additional tolerance to cover the 0.83--0.88 interval given the strong structural prior. This strategy achieves 98.97\% recall (579/585) at candidate generation and provides additional evidence that the selected Stage-2 calibration is reasonably robust.

\subsection{Missed Pairs Analysis}

The 6 missed gold pairs (1.03\%) all have \texttt{kg\_score=NULL}, 
indicating missing document embeddings rather than threshold 
failures. These correspond to two FDA documents (P990071, P990054) 
lacking vector representations in the original dataset.

\section{Cross-Encoder Evaluation Details}
\label{app:reranker}

\subsection{Evaluation Protocol}

We evaluate cross-encoder models on 2,672 samples: 672 gold positives and 2,000 hard negatives (high-similarity non-gold pairs). For each model, we compute:
\begin{itemize}[nosep, leftmargin=*]
    \item \textbf{ROC-AUC}: Area under receiver operating characteristic curve
    \item \textbf{PR-AUC}: Area under precision-recall curve
    \item \textbf{Best F1}: Maximum F1 score across thresholds
    \item \textbf{Latency}: Average inference time per pair (ms)
\end{itemize}

\subsection{Full Results}

Table~\ref{tab:reranker_full} presents complete cross-encoder comparison results.

\begin{table}[h]
\centering
\small 
\caption{Complete cross-encoder comparison (2,672 samples).}
\label{tab:reranker_full}
\resizebox{\linewidth}{!}{
\begin{tabular}{lcccc}
\toprule
\textbf{Model} & \textbf{ROC-AUC} & \textbf{PR-AUC} & \textbf{F1} & \textbf{ms/pair} \\
\midrule
\multicolumn{5}{l}{\textit{Traditional Cross-Encoders}} \\
MiniLM-L12 & .795 & .508 & .590 & 3.2 \\
TinyBERT-L6 & .772 & .609 & .583 & 4.2 \\
MiniLM-L6 & .736 & .513 & .521 & 2.8 \\
BGE-reranker-large & .748 & .536 & .556 & 8.4 \\
BGE-reranker-base & .697 & .512 & .502 & 4.9 \\
BGE-M3$^\dagger$ & .760 & .602 & .574 & 8.4 \\
GTE-ModernBERT & .720 & .497 & .523 & 17.0 \\
\midrule
\multicolumn{5}{l}{\textit{LLM-based Rerankers}} \\
E2Rank-4B & .547 & .302 & .409 & 42.1 \\
ERank-4B & .519 & .302 & .403 & 33.3 \\
mxbai-rerank-large & .527 & .276 & .405 & 16.8 \\
mxbai-rerank-base & .413 & .213 & .403 & 7.9 \\
Jina-Reranker-v3 & .444 & .242 & .404 & 26.1 \\
Qwen3-Reranker-4B & .401 & .203 & .402 & 41.1 \\
\bottomrule
\end{tabular}
}
\begin{tablenotes}
\small
\item $^\dagger$ Deployed model.
\end{tablenotes}

\end{table}

\subsection{Statistical Significance}

We perform bootstrap significance tests (1000 iterations) comparing all models against MiniLM-L12. All differences are statistically significant ($p < 0.001$).

\subsection{Model Selection Rationale}

Despite MiniLM-L12 achieving the highest ROC-AUC (0.795), we deploy BGE-M3 for production due to its 1024-token context window. Patent abstracts frequently exceed 512 tokens; truncation degrades performance on long documents. BGE-M3's slight ROC-AUC reduction (0.760) is offset by improved handling of full-length abstracts.

\section{Gold Standard Construction}
\label{app:gold}

\subsection{Data Sources}

We construct the gold standard by exhaustively searching public disclosures for all 434 cardiovascular devices:

\begin{enumerate}[nosep]
    \item \textbf{Patent litigation filings}: PACER database, ITC Section 337 investigations
    \item \textbf{Virtual patent marking}: Manufacturer websites listing patents covering specific products
    \item \textbf{SEC filings}: 10-K risk factors, 8-K material events mentioning IP
    \item \textbf{Investor presentations}: Quarterly earnings calls, analyst day materials
    \item \textbf{FDA submissions}: PMA summary documents citing prior art
\end{enumerate}

\subsection{Coverage Statistics}

\begin{itemize}[nosep, leftmargin=*]
    \item Devices searched: 434
    \item Devices with disclosed patents: 88 (20.3\%)
    \item Total verified pairs: 585
    \item Patents per device: median = 5, max = 83
\end{itemize}

The low disclosure rate (20.3\%) reflects strategic corporate practices---companies selectively disclose patents for litigation or marketing purposes rather than comprehensive IP mapping.

\subsection{Quality Assurance}

All pairs are derived from legally binding corporate disclosures (patent litigation filings, SEC regulatory submissions, official virtual patent marking pages). These sources carry legal accountability, providing high-confidence ground truth without requiring additional manual verification.
\section{Case Studies}
\label{app:cases}

\subsection{Case A: Recall Root-Cause Analysis}

The Biosense Webster Cardiac Ablation Catheter (PMA P030031) experienced multiple FDA recalls citing ``bi-directional navigation catheter irrigation path'' failures. Our knowledge graph links this device to four patents:
\begin{itemize}[nosep, leftmargin=*]
    \item US7377906: ``Steering mechanism for bi-directional catheter'' (2008)
    \item US7591799: ``Bi-directional catheter steering control'' (2009)
    \item US8021327: ``Irrigated bi-directional catheter'' (2011)
    \item US8348888: ``Multi-directional steering apparatus'' (2013)
\end{itemize}

The recall explicitly implicates the subsystem these patents protect. This enables \textit{proactive surveillance}: querying devices sharing this patent family identifies additional inspection candidates before failures occur.

\subsection{Case B: M\&A-Driven IP Discovery}

The Ovation Stent Graft System (PMA P120006) was originally manufactured by TriVascular Technologies. Our gold standard includes 82 associated patents. Following Endologix's 2016 acquisition (\$211M), many patents transferred ownership while the FDA approval remained under the TriVascular name.

Our entity matching identifies persistent technical concepts (``endovascular graft,'' ``polymer fill,'' ``low-profile delivery'') across corporate boundaries, automatically surfacing M\&A-driven IP relationships without requiring manual corporate genealogy tracking.

\subsection{Case C: Technology Trajectory Mapping}

For Shockwave Medical's Intravascular Lithotripsy (IVL) System (PMA P200039), we recover 19 linked patents spanning 2014--2018:
\begin{itemize}[nosep, leftmargin=*]
    \item 2014: Foundational electrode designs (US8728091)
    \item 2015: Energy delivery control mechanisms (US9011463)
    \item 2016--2017: Catheter integration patents
    \item 2018: Multi-source architecture (US10039561)
\end{itemize}

This trajectory enables: (1) patent expiration forecasting for competitive intelligence, (2) R\&D gap analysis identifying uncovered technical domains, and (3) prior art mapping for freedom-to-operate assessments.

\section{Error Analysis}
\label{erroranalysis}
The 49 unrecovered gold pairs (8.4\%) exhibit: average similarity 0.848 (below 0.92 immunity threshold), average fusion probability 0.294, and 88\% have company match but lack confirming signals. Expert audit categorizes:
\begin{itemize}[nosep, leftmargin=*]
    \item \textbf{Weak links (60.9\%)}: Tangentially related patents disclosed for legal completeness
    \item \textbf{Genuine failures (30.4\%)}: Semantic drift or part-whole mismatch
    \item \textbf{Annotation errors (8.7\%)}: Incorrect gold labels
\end{itemize}
The high proportion of weak links suggests our conservative metrics underestimate true recall.

\section{Computational Cost}
\label{app:compute}
We report the computational resources required by our coarse-to-fine pipeline to support full-corpus processing. Table~\ref{tab:compute} summarizes the end-to-end cost for entity extraction and embedding-based retrieval, which together constitute the dominant computational components of the system.

Entity extraction is executed once over the full patent and medical device corpus using parallel CPU processing. Despite its computational intensity, the task is embarrassingly parallel and completes within approximately 72 hours on 40 CPU workers, incurring a total cost of around CNY 1,600. The 72-hour entity extraction is a one-time initialization cost over the 698K-document corpus; subsequent per-device inference completes in minutes.

Embedding computation and associated ablation experiments are conducted on a single A800 GPU and complete within 8 hours at a marginal cost of approximately CNY 100. Notably, subsequent stages operate only on the reduced candidate set produced by the earlier stages, avoiding full pairwise cross-encoder evaluation over the entire corpus.

In total, the full pipeline completes in under 80 hours with an estimated cost of CNY 1,700, making full-corpus evaluation tractable in contrast to exhaustive cross-encoder scoring, which would require hundreds of millions of comparisons and incur prohibitive computational and API costs.

\begin{table}[h]
\centering
\small
\caption{Computational cost breakdown. \textit{The full pipeline is computationally tractable, avoiding exhaustive cross-encoder evaluation over hundreds of millions of pairs.}}
\label{tab:compute}
\resizebox{\linewidth}{!}{
\begin{tabular}{lccc}
\toprule
\textbf{Stage} & \textbf{Hardware} & \textbf{Time} & \textbf{Cost (CNY)} \\
\midrule
Entity extraction (698K docs) & 40× CPU parallel & 72 hrs & $\sim$1,600 \\
Embedding + Ablations & A800 GPU & 8 hrs & $\sim$100 \\
\midrule
\textbf{Total} & --- & $\sim$80 hrs & $\sim$1,700 \\
\bottomrule
\end{tabular}
}
\end{table}

\end{document}